\documentclass[8.5pt,twoside,twocolumn]{article}
\usepackage[super,sort&compress,comma]{natbib}
\usepackage{mhchem}
\usepackage{times,mathptmx}
\usepackage{sectsty}
\usepackage{balance}

\usepackage{graphicx} 
\usepackage{lastpage}
\usepackage[format=plain,justification=raggedright,singlelinecheck=false,font=small,labelfont=bf,labelsep=space]{caption}
\usepackage{fancyhdr}
\begin{document}

\thispagestyle{plain}
\fancypagestyle{plain}{
\fancyhead[L]{\includegraphics[height=8pt]{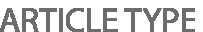}}
\fancyhead[C]{\hspace{-1cm}\includegraphics[height=20pt]{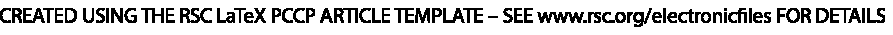}}
\fancyhead[R]{\includegraphics[height=10pt]{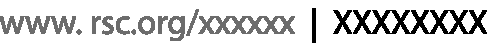}\vspace{-0.2cm}}
\renewcommand{\headrulewidth}{1pt}}
\renewcommand{\thefootnote}{\fnsymbol{footnote}}
\renewcommand\footnoterule{\vspace*{1pt}%
\hrule width 3.4in height 0.4pt \vspace*{5pt}}
\setcounter{secnumdepth}{5}

\makeatletter
\def\subsubsection{\@startsection{subsubsection}{3}{10pt}{-1.25ex plus -1ex minus -.1ex}{0ex plus 0ex}{\normalsize\bf}}
\def\paragraph{\@startsection{paragraph}{4}{10pt}{-1.25ex plus -1ex minus -.1ex}{0ex plus 0ex}{\normalsize\textit}}
\renewcommand\@biblabel[1]{#1}
\renewcommand\@makefntext[1]%
{\noindent\makebox[0pt][r]{\@thefnmark\,}#1}
\makeatother
\renewcommand{\figurename}{\small{Fig.}~}
\sectionfont{\large}
\subsectionfont{\normalsize}

\fancyfoot{}
\fancyfoot[LO,RE]{\vspace{-7pt}\includegraphics[height=9pt]{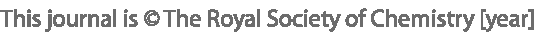}}
\fancyfoot[CO]{\vspace{-7.2pt}\hspace{12.2cm}\includegraphics{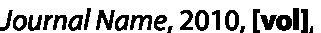}}
\fancyfoot[CE]{\vspace{-7.5pt}\hspace{-13.5cm}\includegraphics{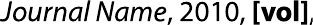}}
\fancyfoot[RO]{\footnotesize{\sffamily{1--\pageref{LastPage} ~\textbar  \hspace{2pt}\thepage}}}
\fancyfoot[LE]{\footnotesize{\sffamily{\thepage~\textbar\hspace{3.45cm} 1--\pageref{LastPage}}}}
\fancyhead{}
\renewcommand{\headrulewidth}{1pt}
\renewcommand{\footrulewidth}{1pt}
\setlength{\arrayrulewidth}{1pt}
\setlength{\columnsep}{6.5mm}
\setlength\bibsep{1pt}

\twocolumn[
  \begin{@twocolumnfalse}
\noindent\LARGE{\textbf{Magnetism of sodium superoxide}}
\vspace{0.6cm}

\noindent\large{\textbf{I. V. Solovyev,$^{\ast}$\textit{$^{a,b}$} Z. V. Pchelkina,\textit{$^{b,c}$} and
V. V. Mazurenko\textit{$^{b}$}}}\vspace{0.5cm}

\noindent\textit{\small{\textbf{Received Xth XXXXXXXXXX 20XX, Accepted Xth XXXXXXXXX 20XX\newline
First published on the web Xth XXXXXXXXXX 200X}}}

\noindent \textbf{\small{DOI: 10.1039/b000000x}}
\vspace{0.6cm}

\noindent \normalsize{By combining first-principles electronic-structure calculations with the model Hamiltonian approach,
we systematically study the magnetic properties of sodium superoxide (\ce{NaO2}),
originating from interacting superoxide molecules.
We show that \ce{NaO2} exhibits a rich variety of magnetic properties,
which are controlled by
relative alignment of the superoxide molecules as well as the state of partially filled
antibonding molecular $\pi_g$-orbitals.
The orbital degeneracy and disorder in the high-temperature pyrite phase
gives rise to weak isotropic antiferromagnetic (AFM) interactions between the molecules.
The transition to the low-temperature marcasite phase lifts the degeneracy, leading to
the orbital order and
formation of
the quasi-one-dimensional AFM spin chains.
Both tendencies are consistent with the behavior of experimental magnetic susceptibility data.
Furthermore, we evaluate the magnetic transition temperature and type of the long-range magnetic order
in the marcasite phase. We argue that this magnetic order depends on the behavior of weak
isotropic as well as anisotropic and Dzyaloshinskii-Moriya exchange interactions
between the molecules.
Finally, we predict the existence of a multiferroic phase, where the inversion
symmetry is broken by the long-range magnetic order, giving rise to substantial ferroelectric polarization.}
\vspace{0.5cm}
 \end{@twocolumnfalse}
  ]



\footnotetext{\textit{$^{a}$~Computational Materials Science Unit,
National Institute for Materials Science, 1-2-1 Sengen, Tsukuba,
Ibaraki 305-0047, Japan. Fax: 81 29 859 2601; Tel: 81 29 859 2619; E-mail: SOLOVYEV.Igor@nims.go.jp}}
\footnotetext{\textit{$^{b}$~Department of Theoretical Physics and Applied Mathematics, Ural Federal University,
Mira str. 19, 620002 Ekaterinburg, Russia.}}
\footnotetext{\textit{$^{c}$~Institute of Metal Physics, Russian Academy of
Sciences-Ural Division, 620990 Ekaterinburg, Russia.}}



\section{\label{sec:intro} Introduction}
Sodium superoxide (\ce{NaO2}) exhibits a rich variety of properties, which make it an interesting example
of multifunctional materials.
Being traditionally used as one of the components of oxygen regeneration devices, it emerged recently as a promising material for
rechargeable room-temperature batteries.\cite{batteries}

  Particularly interesting are the magnetic properties of \ce{NaO2}, which were actively studied in 1970s, together
with other alkali superoxides (or hyperoxides).\cite{KanzigLabhart,Helvetica1976,MahantiSSC,MahantiPRB,magnetogyrationPRL,magnetogyration,Labhart1979}
Despite some exoticism, these systems may be regarded as alternative magnetic materials, which are built without
traditional transition-metal or rare-earth elements, and where the source of the magnetism is the molecular
superoxide complexes O$_2^-$, arranged in a periodic lattice. Such a behavior is rather unique and related to the
Hund's rule effects in the $p$-electron shell of oxygen molecules, which tend to form a high-spin
triplet state.\cite{LandauLifshitz} Since O$_2^-$ is
an aspherical object, a particular attention was paid to exploration of principally new effects,
associated with the rotational degrees of freedoms of the superoxide molecules.
One of such effects is the magnetogyration -- a new type of magnetoelastic coupling between localized magnetic moments
and rotations of the molecular groups.\cite{magnetogyrationPRL,magnetogyration}

  Nevertheless, one very important aspect of molecular complexes was yet overlooked in the early studies of alkali superoxides.
In comparison with the regular oxygen molecule, O$_2^-$ acquires an extra electron in the degenerate $\pi_g$-shell,
which is composed from antibonding molecular orbitals of the $p_x$- and $p_y$-type. Thus, the
properties of superoxides crucially depend on where this electron will reside. This may depend on several factors.
For instance, because of the Jahn-Teller theorem, the degeneracy of the $\pi_g$-orbitals should be lifted
in the ground state, that typically leads to the long-range orbital ordering.
Nevertheless, the direction for lifting the degeneracy depends on the
electron-electron interactions, relativistic spin-orbit (SO) interaction, lattice distortions and --
in this particular case -- the rotations of the superoxide complexes. The magnetic properties of superoxides
will replicate the orbital ordering. In the high-temperature phase, however,
the Jahn-Teller theorem is no longer applicable and
the character of inter-molecular
magnetic interactions can be specified by the orbital disorder in a degenerate state. Such orbital effects are well known for the
transition-metal oxides, where the active orbitals are the atomic $d$-orbitals.\cite{KugelKhomskii,TokuraNagaosa}
The new aspect of superoxides is that the same type of the effects
can be realized on the molecular $\pi_g$-orbitals,
which may bring some new functionalities into the canonical problem of magnetic interactions, due to
specific geometry and antibonding character of these orbitals.
Thus, the recent wave of interest in the alkali superoxides is related to the exploration of the orbital effects
and their impact on the magnetic properties.\cite{NJP08,Kovacik,Kim,Ylvisaker,Nandy,Wohlfeld}
Another important direction is the search of the new type of the ferromagnetic and, possibly, half-metallic materials
on the basis of $p$-elements.\cite{Attema05,RiyadiCM}

  In this work, we will present a comprehensive theoretical analysis of the magnetic properties of \ce{NaO2}.
Amongst alkali superoxides, \ce{NaO2} is one of the most experimentally studied compounds. Particularly,
it is the only compound for which details of the lattice distortion and rotations of the superoxide molecules
have been reported for the low-temperature marcasite phase.\cite{Helvetica1976,Giriyapura}
We will show that \ce{NaO2} is an exciting magnetic material, which bears many similarities with more
traditional transition-metal oxides. The magnetic properties of the high-temperature pyrite phase are
specified by the orbital disorder, which explains isotropic and weakly antiferromagnetic (AFM) character of
inter-molecular interactions. On the contrary, the orthorhombic distortion in the low-temperature marcasite phase
sets up an orbital order, which results in the quasi-one-dimensional character of inter-molecular interactions.
Moreover, the magnetic interactions in this phase are featured by frustration effects. These two factors
explain relatively low magnetic transition temperature of \ce{NaO2}.
Fine details of the magnetic structure are controlled by the SO-related anisotropic
and Dzyaloshinskii-Moriya interactions, which lift the degeneracy
of the ground state.
Moreover, we predict the existence of the multiferroic phase of \ce{NaO2},
where the inversion symmetry is broken by the long-range magnetic order due to the frustration.

\section{\label{sec:method} Method}
\begin{figure}[h]
\centering
  \includegraphics[width=8.3cm]{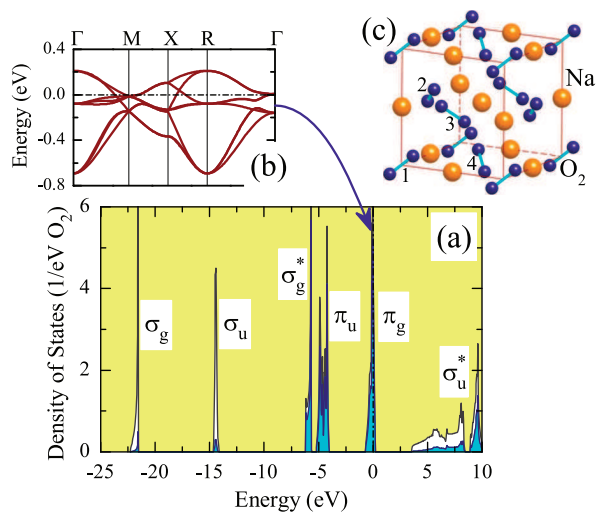}
  \caption{(a) Total and partial densities of states for the pyrite phase of \ce{NaO2}
  in the local-density approximation. Shaded areas show contributions of the oxygen $2p$ states.
  The positions of the main bands are indicated by symbols. The Fermi level is at zero energy (shown by dot-dashed line).
  (b) Energy dispersion of the $\pi_g$-band, located near the Fermi level. Notations of the high-symmetry points
  of the Brillouin zone are taken from the book of \citeauthor{BradlayCracknell}.\cite{BradlayCracknell} (c) Crystal structure of the pyrite phase.}
  \label{fig:pyrite}
\end{figure}

According to the first-principles electronic structure calculations in the local density approximation (LDA),
the strongest hybridization in \ce{NaO2} occurs within the superoxide complexes, which leads to the formation of the
molecular levels. The characteristic splitting between these levels is several eV.
The hybridization between molecules is significantly weaker and leads to the formation of the molecular bands.
The characteristic bandwidth is of the order of one eV. Thus, the bands do not overlap with each other and
can be identified by using the same notations as for the isolated molecular levels. Typical
example of the LDA band structure for the pyrite phase is shown in Figure~\ref{fig:pyrite}.

  LDA is known to be a bad approximation for the description of Coulomb correlations in the narrow-band compounds.
For these purposes, it is essential to go beyond LDA. At the same time, we do not need to know all details of the electronic structure:
since the magnetic properties of \ce{NaO2} are solely related to the behavior of the $\pi_g$-bands, we can concentrate on
the description of only this group of states, by constructing an effective low-energy (Hubbard-type) model and including the effect of
other states to the definition of parameters of the low-energy model. For these purposes, we construct the Wannier basis
for the $\pi_g$-band, using the projector-operator technique and isolated molecular orbitals as the trial functions.\cite{WannierRevModPhys}
Then, matrix elements of LDA Hamiltonian in the Wannier basis give us one-electron part of the model: namely,
the site-diagonal elements describe the crystal-field splitting of the molecular $\pi_g$-orbitals, while the
off-diagonal elements have a meaning of transfer integrals (or the kinetic hoppings between different molecular sites).
The effective Coulomb interactions in the $\pi_g$-band should take into account the screening by other bands.
This screening is calculated by combining constraint-LDA and the random-phase approximation (RPA).
This procedure was previously applied to \ce{KO2},\cite{NJP08} which is similar to \ce{NaO2}, except for the crystal structure.
The details can be also found in the review article.\cite{review2008}

\section{\label{sec:results} Results and Discussions}

\subsection{\label{sec:pyrite} Pyrite phase}
\subsubsection{Structural properties}~The pyrite phase of \ce{NaO2} is realized above 196 K. It has four superoxide ions in the
primitive cell, which are located at $(0,0,0)$, $(0,\frac{1}{2},\frac{1}{2})$, $(\frac{1}{2},0,\frac{1}{2})$, and $(\frac{1}{2},\frac{1}{2},0)$,
respectively,
in units of the cubic lattice parameter $a$.  Each molecule is aligned along
one of the four body diagonals of the cube:
$[1,\overline{1},1]$, $[1,1,\overline{1}]$, $[\overline{1},1,1]$, or
$[1,1,1]$.
Below $223$ K, these molecular axes are
ordered as explained in Figure~\ref{fig:pyrite}c.
The corresponding space group is $Pa\overline{3}$. It is believed that above 223 K,
the molecular axes undergo some hindered rotations, leading to an averaged NaCl-type structure
with the space group $Fm\overline{3}m$. Nevertheless, the diffuse X-ray scattering indicates that, even above 223 K,
the distribution over four possible molecular axes is not entirely random, and the existing between them correlations
correspond to the local order of the pyrite type.\cite{KanzigLabhart}
Since the physically relevant mechanism, responsible for the
inter-molecular exchange interactions in alkali superoxides is the superexchange (SE),\cite{NJP08} which is a local process
and depends only on transfer integrals between neighboring molecules,\cite{PWAnderson,KugelKhomskii}
we do need to know all details of the orientational disorder and
assume the ordered $Pa\overline{3}$ structure in all
temperature range above 196 K. We use experimental parameters of the crystal structure of \ce{NaO2},
reported in ref. \citenum{Helvetica1976}.

\subsubsection{Parameters of electronic model}~The electronic structure of the pyrite phase of \ce{NaO2} in LDA is explained in Figure~\ref{fig:pyrite}.
As was discussed in Section~\ref{sec:method}, our first goal is the construction of an effective Hubbard-type
model for the narrow oxygen $\pi_g$-band, located near the Fermi level.
Due to the high $Pa\overline{3}$ symmetry, the molecular $\pi_g$-orbitals remain degenerate and there is
no crystal-field splitting. Therefore,
the one-electron part of the model
will include only the transfer integrals, connecting different O$_2^-$ sites.
For each pair of the O$_2^-$ molecules, located at $i$ and $j$, the transfer integrals between
molecular $\pi_g$-orbitals are the $2\times2$ real matrices $\hat{t}_{ij}$.
The form of these matrices depends on the relative position of the sites $i$ and $j$, and the choice of the
local coordinate frame at each of these sites.\cite{Note-1}
However, for the purposes of this section, it is sufficient to know only the averaged transfer integrals
$\bar{t}_{ij} \equiv \| \hat{t}_{ij} \|_F$, expressed in terms of Frobenius
matrix norm $\| \hat{t}_{ij} \|_F = \sqrt{{\rm Tr}(\hat{t}_{ij}^{\phantom{T}}\hat{t}_{ij}^T) }$, where
${\rm Tr}$ is the trace over orbital indices.
For the nearest-neighbor (NN) sites, all $\bar{t}_{ij}$ are the same ($\bar{t}_{ij} \equiv \bar{t}$)
and do not depend on the coordinate frame. Using numerical values of transfer integrals $\hat{t}_{ij}$,
$\bar{t}$ can be evaluated as $\bar{t} = 68$ meV. As a test, the bandwidth for the face-centered
cubic lattice can be estimated as $12 \bar{t} = 816$ meV, which is well consistent with results
of LDA calculations for the $\pi_g$-band (Figure~\ref{fig:pyrite}).

  For the $Pa\overline{3}$ symmetry, the screened intra-molecular interactions between $\pi_g$-electrons can be described
in terms of two
independent Kanamori parameters:\cite{Kanamori} the intra-orbital Coulomb repulsion
$U$ and the intra-molecular (Hund's) exchange $J_{\rm H}$.
The third parameter, the so-called inter-orbital Coulomb repulsion $U'$,
is related with the former two by the identity $U' = U - 2J_{\rm H}$. The direct calculations, using the combined
constraint-LDA plus RPA approach for the screening,\cite{NJP08,review2008} yield
$U = 3.65$ eV and $J_{\rm H} = 0.61$ eV.

\subsubsection{Spin Hamiltonian and magnetic properties}~Now, we are ready to construct the
isotropic
(Heisenberg-type) spin Hamiltonian,
\begin{equation}
{\cal H}_H = -\sum_{i>j} J_{ij} {\bf S}_i {\bf S}_j,
\label{eqn:spinH}
\end{equation}
and evaluate parameters of this Hamiltonian for the pyrite phase,
using the theory of SE interactions.\cite{PWAnderson,KugelKhomskii}
In these notations, $S=1/2$ and the summation runs over inequivalent pairs of molecular sites.
The SE theory is basically the second order perturbation theory with respect to the transfer integrals
in the bond $i$-$j$, where one should
evaluate the energy gain (${\cal T}_{ij}$) caused by the virtual hoppings from the space of occupied states at the site $i$ to the
subspace of unoccupied states at the site $j$ (and vice versa) and to do so for the
ferromagnetic (FM, $\uparrow \uparrow$) and AFM ($\uparrow \downarrow$) configurations of spins.
Then, the exchange coupling $J_{ij}$ is obtained by mapping these energies onto the spin Hamiltonian (\ref{eqn:spinH}),
which yields $J_{ij} = ({\cal T}_{ij}^{\uparrow \downarrow}-{\cal T}_{ij}^{\uparrow \uparrow})/(2S^2)$.

  In the O$_2^-$ molecule, there are three electrons, residing on four molecular spin-orbitals of the $\pi_g$-symmetry.
Therefore, there will be two occupied orbitals with the majority ($\uparrow$) spin and
one occupied orbital with the minority ($\downarrow$) spin.
Very generally, the latter (electronic) orbital can be presented in the form:\cite{LandauLifshitz}
\[
| \psi_i^e \rangle = \cos \frac{\vartheta_i}{2} e^{i\varphi/2} | p_x \rangle + \sin \frac{\vartheta_i}{2} e^{-i\varphi/2} | p_y \rangle,
\]
in terms of two $\pi_g$-orbitals of the $p_x$ and $p_y$ symmetry (where the direction of $z$ is taken
along the molecular axis, $0 \leq \vartheta_i \leq \pi$, and $0 \leq \varphi_i \leq 2\pi$).
Then, the unoccupied (hole) orbital should be orthogonal to $| \psi_i^e \rangle$:
\[
| \psi_i^h \rangle = -\sin \frac{\vartheta_i}{2} e^{i\varphi/2} | p_x \rangle + \cos \frac{\vartheta_i}{2} e^{-i\varphi/2} | p_y \rangle.
\]
In the absence of the crystal-field splitting, the angles $\vartheta_i$ and $\varphi_i$
can take arbitrary values. In the ordered state, they should
minimize the total energy of the crystal.\cite{KugelKhomskii} Nevertheless, the pyrite phase exists
only in the high temperature state, where there should be no spin or orbital order. Therefore, in order to consider
spin interactions, $J_{ij}$ should be averaged over $\vartheta_i$ and $\varphi_i$, that would simulate the effect
of the orbital disorder.\cite{KugelKhomskii}

  Let us first evaluate ${\cal T}_{ij}^{\uparrow \uparrow}$. Since the hoppings are permitted only between
orbitals with the same spin, we will have two such contributions, corresponding to the transitions
from $| \psi_i^e \rangle$ to $| \psi_j^h \rangle$ and from $| \psi_j^e \rangle$ to $| \psi_i^h \rangle$.
Therefore, ${\cal T}_{ij}^{\uparrow \uparrow}$ will be given by the following expression:
\[
{\cal T}_{ij}^{\uparrow \uparrow} = -\frac{ |\langle \psi_i^e | \hat{t}_{ij} | \psi_j^h \rangle|^2 + (i \leftrightarrow j)}{U-3J_{\rm H}},
\]
where the denominator is the energy cost for transferring an electron between sites $i$ and $j$ in the ionic limit
(the inter-orbital Coulomb repulsion $U' = U-2J_{\rm H}$ minus the exchange interaction $J_{\rm H}$
in the case of the $\uparrow \uparrow$
configuration of spins). For the $\uparrow \downarrow$ configuration, both states
$| \psi_i^e \rangle$ and $| \psi_i^h \rangle$ with the same spin as $| \psi_j^h \rangle$ will be occupied.
Therefore, ${\cal T}_{ij}^{\uparrow \downarrow}$ will have four contributions:
\[
{\cal T}_{ij}^{\uparrow \downarrow} = -\frac{ |\langle \psi_i^h | \hat{t}_{ij} | \psi_j^h \rangle|^2 + (i \leftrightarrow j)}{U}
-\frac{ |\langle \psi_i^e | \hat{t}_{ij} | \psi_j^h \rangle|^2 + (i \leftrightarrow j)}{U-2J_{\rm H}}.
\]
The energy cost for the first process is $U$ (after transferring an electron from the site $i$, two remaining electrons
occupy the same type of orbitals) and, for the second process, it is $U' = U-2J_{\rm H}$ (remaining electrons occupy
orbitals of different type). Note that there is no exchange
interaction $J_{\rm H}$ between states with different spins.

  Then, we average ${\cal T}_{ij}^{\uparrow \uparrow}$ and
${\cal T}_{ij}^{\uparrow \downarrow}$ over the angles $\vartheta_i$, $\varphi_i$, $\vartheta_j$, and $\varphi_j$,
and treat these angles as independent parameters. This yields
$|\langle \psi_i^h | \hat{t}_{ij} | \psi_j^h \rangle|^2 = |\langle \psi_i^e | \hat{t}_{ij} | \psi_j^h \rangle|^2 = \bar{t}^2/4$.\cite{Note-2}
Then, the NN exchange coupling can be estimated as
\[
J = \bar{t}^2 \left(
\frac{1}{U-3J_{\rm H}} - \frac{1}{U-2J_{\rm H}} - \frac{1}{U}
\right).
\]
Thus, the sign of $J$ is solely determined by the ratio $U/J_{\rm H}$, and for
$U>(3+\sqrt{3})J_{\rm H}$ (which is satisfied for the actual values of $U$ and $J_{\rm H}$),
the NN interaction is AFM. Using numerical values of $U$, $J_{\rm H}$, and $\bar{t}$,
reported above, $J$ can be estimated as $J = -0.63$ meV.
Then, the Curie-Weiss temperature $\theta_{\rm CW} = zJS(S+1)/3 k_{\rm B}$
($z=12$ being the coordination number)
is $\theta_{\rm CW} = -22$ K, which is consistent with the experimental temperature
of $-31$ K.\cite{KanzigLabhart}

  We can apply the same strategy to the high-temperature body-centered tetragonal phase of \ce{KO2}, using parameters
of electronic model reported in ref. \citenum{NJP08}. In this case, two strongest magnetic interactions between
first and second NN (located
along the tetragonal
$a$-axis and the body diagonal of the tetragonal cell)
can be estimated as $-1.38$ and $-3.00$ meV, respectively. This yields $\theta_{\rm CW} = -87$ K, which
is in reasonable agreement with the experimental value
of $-44$ K.\cite{KanzigLabhart} This model analysis naturally explains results of numerical calculations,
reported in ref. \citenum{NJP08}.\cite{Note-3}

  Thus, the negative value of $\theta_{\rm CW}$ is the generic feature of
alkali superoxides in the high-temperature regime,\cite{KanzigLabhart,Riyadi,RiyadiThesis} which can be attributed to the orbital disorder.
The positive value of $\theta_{\rm CW} = 33$ K, observed in the narrow temperature range ($196$ K $< T <$ $223$ K) of \ce{NaO2},
may be related to the onset of the spin-orbital coupling.\cite{Note-4}
Nevertheless, in the higher temperature range, the spin and orbital degrees of freedom become independent from each other,
that eventually leads to the
negative value of $\theta_{\rm CW}$.

\subsection{\label{sec:marcasite} Marcasite phase}
\subsubsection{Structural properties}
\begin{figure}[h]
\centering
  \includegraphics[width=8.3cm]{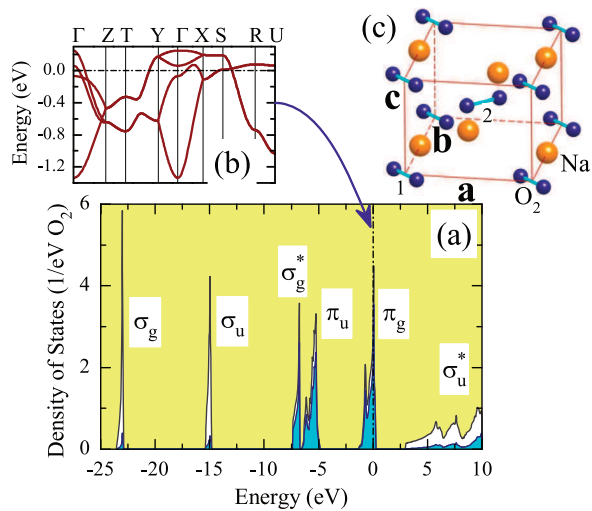}
  \caption{(a) Total and partial densities of states for the marcasite phase of \ce{NaO2}
  in the local-density approximation. Shaded areas show contributions of the oxygen $2p$ states.
  The positions of the main bands are indicated by symbols. The Fermi level is at zero energy (shown by dot-dashed line).
  (b) Energy dispersion of the $\pi_g$-band, located near the Fermi level. Notations of the high-symmetry points
  of the Brillouin zone are taken from the book of \citeauthor{BradlayCracknell}.\cite{BradlayCracknell} (c) Crystal structure of the marcasite phase.}
  \label{fig:marcasite}
\end{figure}

~The marcasite phase of \ce{NaO2} is realized
below $196$ K. It crystallizes
in the orthorhombic $Pnnm$ structure, containing two superoxide ions
in the primitive cell, which are located at $(0,0,0)$ and $(\frac{1}{2},\frac{1}{2},\frac{1}{2})$, respectively,
in units of orthorhombic translations ${\bf a}$, ${\bf b}$, and ${\bf c}$ (see Figure~\ref{fig:marcasite}c).
These ions can be transformed to each other by the glide reflections
$\{ \hat{m}_{\bf a} | {\bf a}/2$$+$${\bf b}/2$$+$${\bf c}/2 \}$ and $\{ \hat{m}_{\bf b} | {\bf a}/2$$+$${\bf b}/2$$+$${\bf c}/2 \}$,
where $\hat{m}_{{\bf a}({\bf b})}$ is the mirror reflection of the orthorhombic ${\bf a}({\bf b})$ axis, and
$({\bf a}/2$$+$${\bf b}/2$$+$${\bf c}/2)$ is the translation along the body diagonal of the
orthorhombic cell.\cite{KanzigLabhart,Helvetica1976}
The magnetic order is realized below $43$ K. However, its type is unknown. In all the calculations we use the
experimental parameters of the crystal structure at $T = 77.3$ K, reported in ref.~\citenum{Helvetica1976}.

\subsubsection{Parameters of electronic model}
\begin{figure}[h]
\centering
  \includegraphics[width=8.3cm]{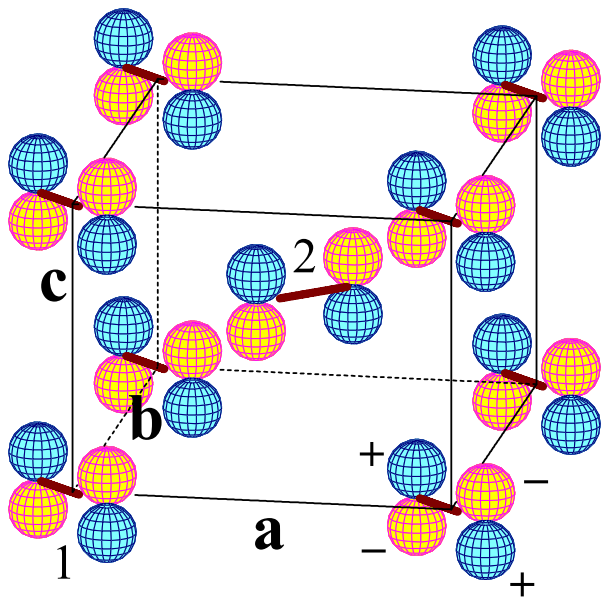}
  \caption{Distribution of unoccupied antibonding molecular
           orbitals of the $p_z$-symmetry (the so-called orbital ordering)
           in the marcasite phase of \ce{NaO2}. Positive and negative lobes
           of the $p_z$-orbitals are shown by different colors. Two sublattices of superoxide ions are denoted
           by indices `1' and `2'.}
  \label{fig:OO}
\end{figure}

~The electronic structure of the marcasite phase of \ce{NaO2}
in LDA is shown in Figure~\ref{fig:marcasite}.
Again, our first goal is the construction of the effective Hubbard-type
model for the oxygen $\pi_g$-band, located near the Fermi level. The new aspect of the marcasite structure is that
the local symmetry is low. The point group includes only the inversion $\hat{I}$, the mirror reflection $\hat{m}_{\bf c}$, and the
$180^\circ$ rotation around the orthorhombic ${\bf c}$-axis ($\hat{C}^2_{\bf c}$). Therefore, the $\pi_g$-levels belong to different
representations and will generally split. More specifically, one of $\pi_g$-orbitals lies in the ${\bf ab}$-plane
(we will call it the $p_{xy}$-orbital, which is a linear combination of the $p_x$- and $p_y$-orbitals) and
another orbital is parallel to the ${\bf c}$-axis (the $p_z$-orbital). We adopt the orbital indices,
where $m=$ $1$ and $2$ correspond to $p_{xy}$ and  $p_z$, respectively. These orbitals are split by the crystal field
and the value of this splitting for the $T = 77$ K structure is about $213$ meV.
Moreover, the $p_z$-orbitals are located higher in energy and periodically ordered, as explained in Figure~\ref{fig:OO}. This orbital
ordering has profound consequences on the behavior of inter-molecular magnetic interactions,\cite{KugelKhomskii}
which will be discussed below.

\begin{figure}[h]
\centering
  \includegraphics[width=8.3cm]{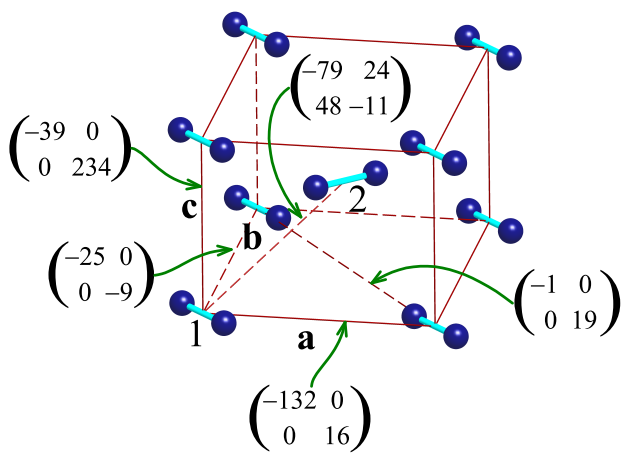}
  \caption{Matrices of transfer integrals $t_{ij}^{mm'}$ (in meV),
           associated with different bonds in the marcasite phase of \ce{NaO2}.
           For each matrix,
           $m(m') = 1$ corresponds to the $p_{xy}$ type of orbitals, and $m(m') = 2$ corresponds
           to the $p_z$ type of orbitals. Two sublattices of superoxide ions are denoted
           by indices `1' and `2'.}
  \label{fig:hoppings}
\end{figure}

  The behavior of the transfer integrals in the basis of the $p_{xy}$- and $p_z$-orbitals is explained in Figure~\ref{fig:hoppings}.
The strongest transfer occurs between $p_z$-orbitals along the orthorhombic
${\bf c}$-axis. Nevertheless,
this is to be expected from the form of the neighboring $p_z$-orbitals, which are directed towards each other.
As we will see below, these transfer integrals are responsible for the strong AFM coupling and formation
of the quasi-one-dimensional AFM chains along ${\bf c}$. The second strongest transfer
occurs between the $p_{xy}$-orbitals along the ${\bf a}$-axis. Nevertheless,
in the marcasite phase, all $p_{xy}$-orbitals are occupied and,
therefore, transfer integrals between them will not contribute to the SE interactions. Somewhat strong transfer
occurs between NN sites of the sublattices $1$ and $2$ along the body diagonals. Other transfer integrals are small. Nevertheless,
as we will see below, they play an important role in the formation of the long-range magnetic order at finite $T$.

  The values of screened intra-orbital Coulomb repulsion $U$ and the exchange interaction $J_{\rm H}$ are
$3.68$ and $0.61$ eV, respectively. In principle, due to the low symmetry, the screening of $U$ will by
slightly different for the orbitals of the $p_{xy}$- and $p_z$-type. However, this difference
is only about 1\% and can be neglected.

\subsubsection{Parameters of spin Hamiltonian}
\begin{figure}[h]
\centering
  \includegraphics[width=8.3cm]{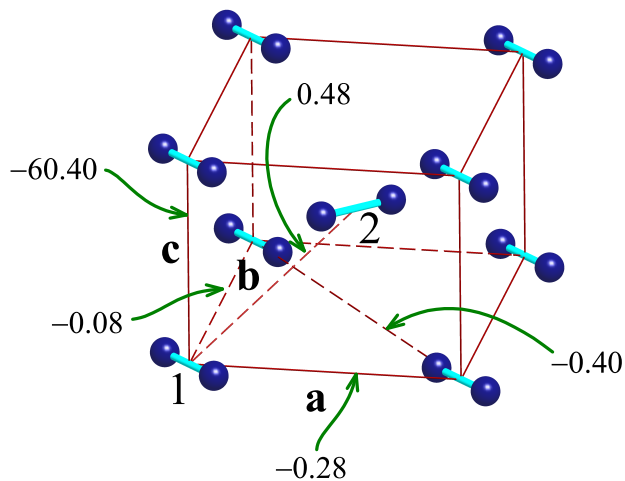}
  \caption{Isotropic exchange interactions (in meV),
           associated with different bonds in the marcasite phase of \ce{NaO2}.
           Two sublattices of superoxide ions are denoted
           by indices `1' and `2'.}
  \label{fig:J}
\end{figure}

~Since all transfer integrals are small in comparison with
the Coulomb repulsion, the parameters of spin Hamiltonian (\ref{eqn:spinH}) can be also evaluated
using the SE theory, and considering all types virtual hoppings into the subspace of unoccupied $p_z$-states.\cite{PWAnderson,KugelKhomskii}
The behavior of obtained parameters is explained in Figure~\ref{fig:J}.
Within each magnetic sublattice, all transfer integrals are diagonal with respect to the orbital indices.
Then, since all $p_{xy}$ orbitals are occupied,
only $p_z$-orbitals will contribute
to the SE interactions.
For the sublattice $1$, the transfer integrals between these
$p_z$-orbitals are (in meV): $t_{\bf a} = 16$, $t_{\bf b} = -$$9$, $t_{\bf c} = 234$, and
$t_{{\bf a}-{\bf b}} = 19$, operating along to the orthorhombic axes ${\bf a}$, ${\bf b}$, and ${\bf c}$, and the
face diagonal ${\bf a}$$-$${\bf b}$, respectively (see Figure~\ref{fig:hoppings}).
The transfer integrals $t_{{\bf a}+{\bf b}}$ along
another face diagonal are small.
This seems to be natural.
Along ${\bf a}$$-$${\bf b}$, the molecular axes are oriented towards each other, leading to larger overlap
between the molecular orbitals. Furthermore, the ${\bf a}$-axis is about 20 \% shorter than ${\bf b}$,
and the molecular axes are additionally canted towards ${\bf a}$. This will lead to a larger
overlap along ${\bf a}$. Nevertheless, we would like to emphasize that all these effects are
sensitive to the details of the crystal structure and the additional rotations of the O$^-_2$ molecules
may change the situation. Some effects of these rotations will be discussed in Section~\ref{sec:TN}.

  For the sublattice $2$,
one should interchange $t_{{\bf a}-{\bf b}}$ and $t_{{\bf a}+{\bf b}}$.

  Corresponding parameters of
isotropic exchange interactions for $S=1/2$ can be obtained using the formula
$J_{\bf R} = -4t_{\bf R}^2/U$. Thus,
$J_{\bf R}$ in Figure~\ref{fig:J} simply reflects the behavior of transfer integrals, depicted in Figure~\ref{fig:hoppings}.
The transfer integrals between different sublattices have large off-diagonal elements, which will contribute to both
FM and AFM coupling.
These contributions are inversely proportional to
$(U$$-$$3J_{\rm H})$ and $(U$$-$$2J_{\rm H})$, respectively. Because of $J_{\rm H}$, the FM contributions will prevail.
On the other hand, the transfer integrals
between $p_z$-orbitals, which contribute only to the AFM coupling, are small. This explains the FM character of NN interactions
between different sublattices $J_{12} = 0.48$ meV.

  The isotropic exchange interactions are frustrated. The largest interaction $J_{\bf c}$ will tend to align the spins in each
sublattice antiferromagnetically. Then, each superoxide ion will form an equal number of FM and AFM bonds with the
NN ions of other sublattice. In such a situation, the spins of different sublattices can arbitrarily rotate
relative to each other and the ground state will be infinitely degenerate. The inter-sublattice coupling $J_{12}$
will lift this degeneracy in favor of a classically ordered spin-spiral state. However, since
$|J_{\bf c}| \gg J_{12}$, the effect is small and can be neglected. We believe that a more realistic scenario for
lifting the degeneracy is the exchange striction and (or) anisotropic exchange interactions. The latter will be considered
in Section~\ref{subsec:anisotripic}.

\subsubsection{Magnetic susceptibility}~In this section, we calculate the magnetic susceptibility ($\chi$) of the spin model (\ref{eqn:spinH}),
using realistic parameters derived above. For these purposes we employ the quantum Monte Carlo loop algorithm,\cite{Alet} as
implemented in the ALPS simulation package.\cite{ALPS1,ALPS2} The calculations have been performed using periodic boundary conditions
for the superlattice including 60 unit cells along the ${\bf c}$-axis, and 2 unit cells along the ${\bf a}$- and ${\bf b}$-axes.

\begin{figure}[h]
\centering
  \includegraphics[width=8.3cm]{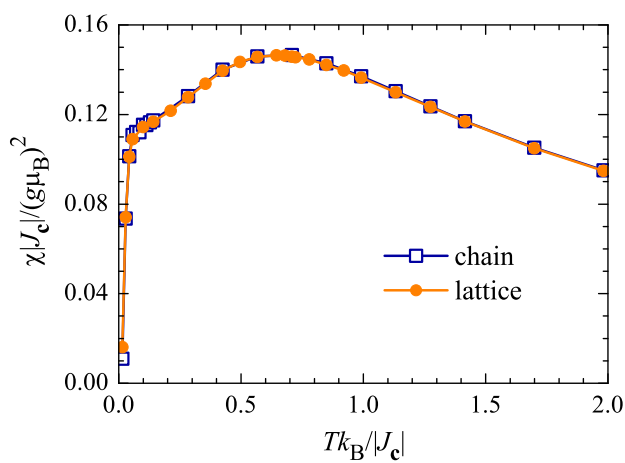}
  \caption{Magnetic susceptibility (per one superoxide ion), calculated for the isolated
  antiferromagnetic chain with the nearest-neighbor exchange interaction $J_{\bf c}$ and the three-dimensional lattice of weakly interacting chains.}
  \label{fig:chi}
\end{figure}

  Results of these calculations are explained in Figure~\ref{fig:chi}.
The shape of $\chi$ is mainly controlled by the intrachain interaction $J_{\bf c}$. Since other interactions are small,
the magnetic susceptibility is practically indistinguishable from that of the isolated AFM chain with the leading
NN coupling $J_{\bf c}$. The latter was discussed in details by Bonner and Fisher in their canonical work.\cite{BonnerFisher}
Namely, $\chi$ has a broad maximum at $T_{\rm max} \approx 0.64 |J_{\bf c}|/k_{\rm B}$.
For $|J_{\bf c}| = 60.4$ meV (see Figure~\ref{fig:J}), it should correspond to $T_{\rm max} \approx 550$ K.
Thus, in the region $43$ K $< T <$ $196$ K,
the magnetic susceptibility is expected to increase.
This finding is consistent with the experimental data.\cite{KanzigLabhart}
It also supports earlier considerations,\cite{MahantiSSC,MahantiPRB} based on the semi-empirical model analysis.
However, the absolute value of susceptibility
at the maximum, $\chi_{\rm max} \approx 0.15 (g \mu_{\rm B})^2/|J_{\bf c}|$,\cite{BonnerFisher} can be estimated as
$\chi_{\rm max} \approx 3\cdot10^{-4}$ cm$^3$/mol (using $g=2$ for the Land\'e $g$-factor). This is about 3-5 times smaller
than the experimental value of susceptibility.\cite{KanzigLabhart} Naively, one would conclude that
such a discrepancy is caused by the overestimation of $|J_{\bf c}|$ (for example, due to the underestimation of the
Coulomb repulsion $U$). However, the situation is not so simple: if $|J_{\bf c}|$ were smaller, $T_{\rm max}$ would be also
smaller and we would have serious difficulties in explaining nearly linear experimental dependence of $\chi$ on $T$
in the temperature interval $43$ K $< T <$ $196$ K. The discrepancy can be partly resolved by using larger value of
$g \approx 2.05 \div 2.10$, due to partial unquenching of the local orbital moment. However, this would increase
theoretical $\chi$ only by less than 10\%. Thus, the discrepancy persists. Apparently, the experimental data themselves
may be influenced by the defects and other extrinsic effects.
Note that the marcasite phase exists only in the finite temperature range. In such a situation, it is not easy to find the
asymptotic behavior and subtract, for instance, the uniform contribution to $\chi$, which may be affected by the sample
surroundings.\cite{RiyadiThesis} We, therefore, propose to carefully reexamine the magnetic susceptibility data.
Otherwise, the quasi-one-dimensional picture for the behavior of magnetic susceptibility cannot explain
simultaneously the linear character and the absolute value of $\chi$.

\subsubsection{\label{sec:TN} Magnetic transition temperature}~As was pointed out above, due to frustrations, the isotropic
exchange interactions alone cannot stabilize the magnetic order between different sublattices. Nevertheless, they do
stabilize the magnetic order in each of the sublattices, although the situation is also not simple and severely
hampered by quasi-one-dimensional character of the problem. If one considers only the strongest AFM
interactions within the chain, according to the Mermin-Wagner theorem,\cite{MerminWagner}
there would be no long range magnetic order at finite $T$. Therefore, the long-range order, if exists, is stabilized by
weak interchain interactions. In this section, we evaluate the transition temperature of this magnetic order.

  First, from the mean-field approximation for the Heisenberg model (\ref{eqn:spinH}),
we find that the classical ground state corresponds to the propagation vector ${\bf Q} = (\pi/a,0,\pi/c)$.
Thus, the magnetic structure is AFM along the orthorhombic ${\bf a}$- and ${\bf c}$-axes, and FM along the ${\bf b}$-axis.
It is favored by relatively strong interactions $J_{\bf a}$ and $J_{{\bf a}-{\bf b}}$ (for the sublattice $1$), and
is disfavored by the weak interaction $J_{\bf b}$ (see Figure~\ref{fig:J}). Then, the N\'{e}el temperature can be
found from the expression:\cite{Schulz,IrkhinKatanin}
\[
k_{\rm B} T_{\rm N} = z_{\perp} | J_{\perp}\ |  A \ln^{1/2} \left( \frac{\Lambda |J_{\bf c}|}{k_{\rm B} T_{\rm N}} \right),
\]
which combines
exact results for the analytically solvable model in one dimension with the mean-field treatment
for the interchain interactions.\cite{Schulz} The numerical calculations
for the single chain yield $A\approx 0.32$ and $\Lambda\approx 5.8$.\cite{Starykh}
In the original formulation,\cite{Schulz,IrkhinKatanin} $J_{\perp}$ is an effective
interchain exchange coupling and $z_{\perp}$ is the coordination number in the ${\bf ab}$-plane.
Therefore, for our purposes, $z_{\perp} | J_{\perp}\ |$ should be replaced by
$2(|J_{\bf a}| + |J_{{\bf a}-{\bf b}}| - |J_{\bf b}|) = 1.2$ meV, which yields $T_{\rm N} \approx 11$ K.
Note that, due to the frustration of magnetic interactions, there will be no mean field
acting on the spins of the sublattice 1 from the sublattice 2 (and vice versa).

  The mean-field theories generally overestimate $T_{\rm N}$. Therefore,
as an alternative possibility, we also considered the random-phase approximation (RPA),\cite{Tyablikov}
\begin{equation}
\frac{1}{k_{\rm B} T_{\rm N}} = \sum_{\bf q} \frac{4}{J({\bf Q}) - J({\bf q})},
\label{eqn:TNRPA}
\end{equation}
which takes into account collective excitations and is expected
to provide a better description for the interchain coupling.\cite{IrkhinKatanin}
In this case, $J({\bf q}) = \sum_{\bf R} J_{\bf R} e^{i {\bf qR}}$ is the Fourier image of $\{ J_{\bf R} \}$.
Then, Eq.~(\ref{eqn:TNRPA}) yields $T_{\rm N} \approx 6$ K, which is consistent with the above mean-field estimate.

  According to the experimental data,\cite{KanzigLabhart} the magnetic order (of unknown type)
is believed to occur below $43$ K, which is substantially higher than our estimate of $T_{\rm N}$.
However, it should be noted that the value of $T_{\rm N}$ is controlled by very small magnetic interactions
between the chains, which may be sensitive to the rotations of the O$^-_2$ molecules. Since we used the
experimental structure parameters measured at $T = 77$ K (i.e., far above $T_{\rm N}$),
it is possible that the additional rotations of O$^-_2$ below $T_{\rm N}$ may change the situation.
In order to check this possibility, we have performed additional calculations, using the experimental
structure parameters at $T = 18$ K, which were recently reported in ref.~\citenum{Giriyapura}.
We have found that $J_{\bf c}$ does not change too much (the new value is $J_{\bf c}=-$$59$ meV).
Other exchange interactions are also relatively small. However, they undergo some changes, which
do affect $T_{\rm N}$. Namely, the strongest interchain interactions for the $T = 18$ K structure
are $J_{{\bf a}-{\bf b}}=-$$0.61$ meV and $J_{{\bf a}+{\bf c}}=-$$0.63$ meV. They stabilize the
classical ground state with ${\bf Q} = (0,\pi/b,\pi/c)$. The corresponding magnetic
transition temperature can be estimated in the RPA as $T_{\rm N} \approx 50$ K. Thus, the discrepancy
with the experimental data can be resolved by taking into account the additional rotations of the
O$^-_2$ molecules below $T_{\rm N}$.

\subsubsection{\label{subsec:anisotripic}Anisotropic interactions}~In this section, we consider
anisotropic exchange interactions, which are caused by the relativistic SO coupling, namely,
the anisotropic symmetric interactions:
\[
{\cal H}_{AS} = \sum_{i>j} {\bf S}_i \hat{\tau}_{ij} {\bf S}_j ,
\]
and the antisymmetric Dzyaloshinskii-Moriya (DM) interactions:\cite{Dzyaloshinskii,Moriya}
\[
{\cal H}_{DM} = \sum_{i>j} {\bf d}_{ij} [ {\bf S}_i \times {\bf S}_j ].
\]
Both types of interactions operate only between different molecular sites, and
due to the Kramers theorem for the spin $1/2$, there will be no single-ion anisotropy.
The parameters $\hat{\tau}_{ij}$ and ${\bf d}_{ij}$
can be also obtained by considering the effects of the SO coupling in the framework of the SE theory.\cite{NJP09}
These interactions are relatively weak. Nevertheless, they play a very important role in
lifting the degeneracy of the magnetic ground state and stabilizing the
inter-sublattice magnetic order. The effect of these interactions can be rationalized in the following way.

  The main contribution to ${\cal H}_{AS}$ comes from the NN interactions along
the ${\bf c}$-axis within sublattices. Corresponding tensor $\hat{\tau}_{\bf c}$ for the sublattice $1$ has the following form (in meV):
\[
\hat{\tau}_{\bf c} = \left(
\begin{array}{ccc}
      -0.04  & \phantom{-}0.47  &  0  \\
       \phantom{-}0.47  & -0.04  &  0  \\
       0     &  0     &  \phantom{-}0.08 \\
\end{array}
\right).
\]
This means that the easy magnetization direction lies in the ${\bf ab}$-plane and is perpendicular to the molecular axes.
As expected, $\hat{\tau}_{\bf c}$ is invariant under the mirror reflection $\hat{m}_{\bf c}$, which
transform the sublattice 1 to itself.
Similar tensor for the sublattice $2$ can be obtained by the mirror reflections $\hat{m}_{\bf a}$ or $\hat{m}_{\bf b}$, which,
in the combination with the translations $({\bf a}$$+$${\bf b}$$+$${\bf c})/2$, transform the sublattice 1
to the sublattice 2.
Other anisotropic symmetric interactions are small and can be neglected. Moreover, since superoxide ions
are located in the inversion centers, there will be no DM interactions operating within each of sublattices.
Thus, the anisotropic symmetric interactions will form a noncollinear magnetic structure where
all spins lie in the ${\bf ab}$-plane, perpendicular to their molecular axes.

  The DM interactions operate between different sublattices.
For example, in the bond 1-2 (see Figure~\ref{fig:J}), the DM vector is
${\bf d}_{12}=(0.06,-0.01,0.01)$ meV. The DM parameters in other NN bonds can be obtained
by applying the symmetry operations of the space group $Pnnm$.\cite{PRL96}
Then, in the classical picture, every spin will experience a force
${\bf f}_i = -\partial {\cal H}_{DM}/\partial {\bf S}_i = \sum_j [ {\bf d}_{ij} \times {\bf S}_j ]$
from its neighboring sites.
Since all spins $\{ {\bf S}_j \}$ are confined in the ${\bf ab}$-plane
and form the ${\bf Q} = (\pi/a,0,\pi/c)$ AFM structure,
it is easy to show that
${\bf f}_i$ should
be parallel to the ${\bf c}$-axis. Thus, the DM interactions will lead to a
small canting of spins off the ${\bf ab}$-plane. Moreover, since
${\bf d}_{21} = -$${\bf d}_{12}$, ${\bf f}_i$ will have opposite
directions at different molecular sites of the unit cell. Therefore, we expect no weak
ferromagnetism in this case.\cite{Dzyaloshinskii,Moriya}

\subsubsection{Multiferroic phase}~
The phenomenon of multiferroicity, or the coupling between magnetism and ferroelectricity,
was intensively studied in transition-metal oxides.\cite{CheongMostovoy,Kimura,Khomskii}
The key moment for obtaining the spontaneous ferroelectric (FE) polarization is to break the
inversion symmetry of the crystal. Therefore, a particular attention is paid to so-called
improper ferroelectricity, where this inversion symmetry is broken by some complex magnetic order.
Such a situation typically occurs in frustrated magnets.
In this section, we will argue that similar behavior is expected for the marcasite
phase of \ce{NaO2}.

\begin{figure}[h]
\centering
  \includegraphics[width=8.3cm]{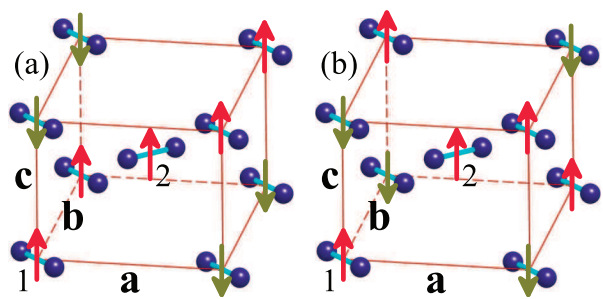}
  \caption{(a) centrosymmetric ${\bf Q} = (\pi/a,0,\pi/c)$
  and (b) noncentrosymmetric ${\bf Q} = (\pi/a,\pi/b,\pi/c)$
  magnetic structures in the marcasite phase of \ce{NaO2}.}
  \label{fig:multiF}
\end{figure}

  Let us consider first the theoretical magnetic ground state, which was obtained from the analysis
inter-molecular magnetic interactions without relativistic spin-orbit (SO) coupling. In this case, each sublattice
forms the ${\bf Q} = (\pi/a,0,\pi/c)$ structure, which is shown in Figure~\ref{fig:multiF}a.
Strictly speaking,
the relative directions of spins in two magnetic sublattices are determined by the SO interactions,
as was argued in the previous section.
However, such fine details of the magnetic structure are not important for the purposes of this section
and here we neglect the effects of the SO coupling. The ${\bf Q} = (\pi/a,0,\pi/c)$ structure can be transformed to
itself by the inversion operation ($\hat{I}$) around any of the superoxide ions. Therefore, the inversion symmetry
is preserved and the ${\bf Q} = (\pi/a,0,\pi/c)$ structure is not ferroelectric.
Similar situation occurs for ${\bf Q} = (0,\pi/b,\pi/c)$.

  In the ${\bf Q} = (\pi/a,\pi/b,\pi/c)$ structure, each magnetic sublattice can be transformed to itself
by combing $\hat{I}$ with the time-reversal operation ($\hat{T}$). This can be clearly seen in Figure~\ref{fig:multiF}b,
by considering the transformations of ions of the sublattice 1 around the inversion center 2. However, the existence of the
symmetry operation $\hat{T}\hat{I}$ would imply that the ion 2, which transforms to itself, is nonmagnetic.
This contradicts to the electronic configuration of O$_2^-$, which has odd number of electrons. Then, the only possibility
to reconcile the ${\bf Q} = (\pi/a,\pi/b,\pi/c)$ type of the magnetic ordering with the electronic configuration of O$_2^-$ is
to break the inversion symmetry. This will produce finite FE polarization ${\bf P}$.
The mechanism is similar to the magnetic inversion symmetry breaking in the $E$-phase of manganites.\cite{PRB12}
The polarization itself can be calculated
using the Berry-phase theory,\cite{KSV,Resta} which was adopted for the effective Hubbard model.\cite{PRB12}
Then, using the wavefunctions, obtained
from the solution of the effective Hubbard model in the Hartree-Fock approximation,
we obtain ${\bf P} = (53,41,-$$248)$ $\mu$C/m$^2$. Thus, the ${\bf Q} = (\pi/a,\pi/b,\pi/c)$ AFM order
eventually
breaks all symmetry operations (not only $\hat{I}$), and ${\bf P}$ has finite values along all three
orthorhombic axes. The absolute value of ${\bf P}$ is more than order of magnitude smaller than
in the $E$-phase of manganites.\cite{PRB12} Nevertheless, this is to be expected because the value of ${\bf P}$
is controlled by the ratio of transfer integrals to the characteristic energy splitting
between molecular $\pi_g$-orbitals (the atomic $e_g$ orbitals in the case of manganites).\cite{PRB13}
The transfer integrals are smaller in superoxides, while the energy splitting is larger
(due to larger Coulomb repulsion). Therefore, $|{\bf P}|$ should be smaller.
Nevertheless, this value of ${\bf P}$ is comparable or even larger than typical values of
FE polarization in many transition-metal oxides with the deformed spin-spiral structure.\cite{Kimura}

\section{Conclusions}~
Using results of first-principles electronic structure calculations
we have systematically studied the magnetic properties of sodium superoxide.
Our basic strategy was to construct a realistic model, describing the behavior of the magnetic degrees of
freedom of \ce{NaO2} in two crystallographic phases, depending on the local environment
and relative orientation of the superoxide molecules. This model provides a transparent picture,
explaining the magnetic properties of \ce{NaO2} in different temperature regimes.
Namely, the high-temperature pyrite phase is characterized by the orbital degeneracy and the
disorder of the
orbital degrees of freedom. This disorder naturally explains the weakly AFM character of
inter-molecular
magnetic interactions, that is seen in the experimental behavior of magnetic susceptibility.
Moreover, we expect that such a behavior should be a generic property of alkali superoxides in the high-temperature state.
The transition to the low-temperature marcasite phase of \ce{NaO2}, which takes place around $196$ K,
is accompanies by the
long-range order of the molecular $\pi_g$-orbitals. This orbital order naturally explains the magnetic properties
of the marcasite phase, namely: (i) The character of isotropic exchange interactions, which is featured by the
formation of the quasi-one-dimensional AFM $S=1/2$ spin chains. This property is again reflected in the behavior of
magnetic susceptibility in the marcasite phase, where $\chi$
rises with the temperature; (ii) Small values of the magnetic transition temperature,
which is related to the one-dimensional character and frustration of isotropic exchange interactions;
(iii) The behavior of anisotropic exchange interactions, which plays an important role in stabilizing
the magnetic order between different magnetic sublattices. Moreover, we predict
the multiferroic behavior of the marcasite phase. The theoretical value of
polarization $|{\bf P}| \sim 250$ $\mu$C/m$^2$ is comparable with those observed in the transition-metal oxides,
which are currently intensively studied in the context of multiferroic applications.

  Thus, the alkali superoxides present an interesting example of magnetic materials on the basis of $p$-elements.
They exhibit fascinating magnetic properties, including the spin-orbital coupled phenomena and
the multiferroic effect. In this sense, the alkali superoxides can be regarded as
molecular $p$-electron analogs of more traditional transition-metal oxides.
The new aspect of alkali superoxides is that all these phenomena are realized on antibonding molecular orbitals of the
superoxide complexes, that opens new functional possibilities for the design and control of the properties.
Another important issue is the rotations of the superoxide molecules -- the problem, which deserves a more
fundamental theoretical analysis.

\section*{Acknowledgements}~
We thank A.~A. Katanin and S.~V. Streltsov for helpful discussions.
This work is partly supported by the grant of the Ministry of
Education and Science of Russia N 14.A18.21.0889.
The work of Z.V.P. is partly supported by the grant RFFI-12-02-31331.




\footnotesize{
\providecommand*{\mcitethebibliography}{\thebibliography}
\csname @ifundefined\endcsname{endmcitethebibliography}
{\let\endmcitethebibliography\endthebibliography}{}

\bibliographystyle{rsc} 

\begin{mcitethebibliography}{51}
\bibitem[Not()]{Note-1}
Note that, due to the $Pa\overline{3}$ symmetry, the local coordinate frame at
  each site can be arbitrarily rotated around the molecular axis. The form of
  $\hat{t}_{ij}$ will depend on these rotations, although the physical
  properties of \ce{NaO2} will not.\relax
\mciteBstWouldAddEndPunctfalse
\mciteSetBstMidEndSepPunct{\mcitedefaultmidpunct}
{}{\mcitedefaultseppunct}\relax
\EndOfBibitem
\bibitem[Not()]{Note-3}
Note that, in ref. \citenum{NJP08}, the parameters $J_{ij}$ were additionally
  multiplied by $S^2=1/4$.\relax
\mciteBstWouldAddEndPunctfalse
\mciteSetBstMidEndSepPunct{\mcitedefaultmidpunct}
{}{\mcitedefaultseppunct}\relax
\EndOfBibitem
\bibitem[Not()]{Note-4}
In fact, a similar scenario was proposed in ref.~\citenum{NJP08} for \ce{KO2}:
  The SO interaction in superoxides, $\xi/k_{\rm B}$, is of the order of $200$
  K. Therefore, around this temperature, the spin and orbital degrees of
  freedom become coupled, and this coupling can lead to FM anisotropic
  interactions between the molecules.\relax
\mciteBstWouldAddEndPunctfalse
\mciteSetBstMidEndSepPunct{\mcitedefaultmidpunct}
{}{\mcitedefaultseppunct}\relax
\EndOfBibitem
\end{mcitethebibliography}
}

\end{document}